\documentclass[letter]{aa} 
\usepackage{graphicx}
\usepackage{float}
\usepackage{indentfirst}
\usepackage{lscape}
\usepackage{longtable}

\usepackage{amsmath}
\usepackage{amssymb}
\usepackage{graphicx}
\usepackage{lastpage}
\usepackage{natbib}
\usepackage{txfonts}

\begin{document}
   \title{First detection of HS$_2$ in a cold dark cloud}

   \author{G. Esplugues
          \inst{1, 2},          
          M. Agúndez
          \inst{3},
          G. Molpeceres
          \inst{3}, 
          B. Tercero
          \inst{1, 2},
          C. Cabezas
          \inst{3},
          N. Marcelino  
          \inst{1, 2},
          R. Fuentetaja
          \inst{3},
          J. Cernicharo
          \inst{3}
         }

   \institute{Observatorio Astron\'omico Nacional (OAN), Alfonso XII, 3, 28014. Madrid. Spain\\
              \email{g.esplugues@oan.es, jose.cernicharo@csic.es}       
        \and
Observatorio de Yebes, IGN, Cerro de la Palera s/n, E-19141 Yebes, Guadalajara, Spain
        \and
Instituto de Física Fundamental, CSIC, Calle Serrano 123, E-28006 Madrid, Spain 
}

  \abstract   
   {We report the first detection of HS$_2$ towards the cold dark cloud TMC-1. This is the first observation of a chemical species containing more than one sulphur atom in this type of sources. The astronomical observations are part of QUIJOTE, a line survey of TMC-1 in the Q band (31-50 GHz). The detection is confirmed by the observation of the fine and hyperfine components of two rotational transitions (2$_{0}$$_{,}$$_{2}$-1$_{0}$$_{,}$$_{1}$ and 3$_{0}$$_{,}$$_{3}$-2$_{0}$$_{,}$$_{2}$). Assuming a rotational temperature of 7 K, we derived an HS$_2$ column density of 5.7$\times$10$^{11}$ cm$^{-2}$, using a local thermodynamic equilibrium model that reproduces the observed spectra. The abundance of HS$_2$ relative to H$_2$ is 5.7$\times$10$^{-11}$, which means that it is about seven times more abundant than its oxygenated counterpart HSO. We also explored the main formation and destruction mechanisms of HS$_2$ using a chemical model, which reproduces the observed abundance of HS$_2$ and indicates that dissociative recombination reactions from the ions H$_2$S$^{+}_{\mathrm{2}}$ and H$_3$S$^{+}_{\mathrm{2}}$ play a major role in forming HS$_2$.}

   \keywords{ISM: abundances - ISM: clouds - ISM: molecules - Radio lines: ISM}
   \titlerunning{First detection of HS$_2$ in a cold dark cloud}
   \authorrunning{G. Esplugues et al.}
   \maketitle

\section{\textbf{Introduction}}

Up to date, over 330 molecules have been detected\footnote{https://cdms.astro.uni-koeln.de/classic/molecules} in the interstellar medium (ISM), from which about 10$\%$ contain sulphur atoms. Ultra-sensitive spectral surveys carried out in the last few years, such as QUIJOTE  \citep[e.g.][]{Cernicharo2021a, Cernicharo2021b} and GOTHAM \citep[e.g.][]{Remijan2025}, have allowed the discovery of about 20$\%$ of all known sulphur species in space. Sulphur (S) is the tenth most abundant element in the Universe and is known to play a significant role in biological systems, with presence in different biomolecules \citep[e.g.][]{Chatterjee2022}. On Earth, S supports the chemoautotrophic and the photosynthetic ways of life \citep[e.g.][]{Clark1981, Narayan2023}. In particular, it acts as a nutrient for plants and animals, since it is a vital component of several biochemical compounds like proteins that form such amino acids, such as methionine \citep[]{Townsend2004}. 
Sulphur is also key to the synthesis of some vitamins, such as C, B1, and B8 \citep[e.g.][]{Raulin1977, Begley1999}.

\begin{table*}
\centering
\caption{Line parameters obtained from Gaussian fits of the detected HS$_2$ lines in TMC-1.}
\begin{tabular}{lllllllll}
\hline 
\hline
         &              & Rest          &  $E$$_{\mathrm{up}}$ & $A$$_{\mathrm{ul}}$ &  v$_{\mathrm{LSR}}$ & $\Delta$v     & $T$$^{\star}_{\mathrm{A}}$  & $\int$$T$$^{\star}_{\mathrm{A}}$dv  \\
Species  & Transition   & Frequency     &  (K)                 & (s$^{-1}$)          & (km s$^{-1}$)       & (km s$^{-1}$) & (mK)                 & (mK km s$^{-1}$)                \\
         &              & (MHz)         &                      &                     &                     &               &                      &                               \\

\hline
\hline 
HS$_2$   & 2$_{0}$$_{,}$$_{2}$-1$_{0}$$_{,}$$_{1}$, $J$ = 5/2-3/2, $F$ = 2-1  &  31440.85  & 2.2 & 1.74e-7 & 5.86$\pm$0.24   & 1.65$\pm$0.20  & 0.60$\pm$0.22  & 1.12$\pm$0.38 \\
HS$_2$   & 2$_{0}$$_{,}$$_{2}$-1$_{0}$$_{,}$$_{1}$, $J$ = 5/2-3/2, $F$ = 3-2  &  31442.42  & 2.2 & 1.95e-7 & 6.10$\pm$0.36   & 1.85$\pm$0.18  & 1.02$\pm$0.28  & 2.02$\pm$0.55 \\
HS$_2$   & 2$_{0}$$_{,}$$_{2}$-1$_{0}$$_{,}$$_{1}$, $J$ = 3/2-1/2, $F$ = 2-1  &  31650.63  & 2.2 & 1.67e-7 & 5.86$\pm$0.21   & 1.20$\pm$0.10  & 0.47$\pm$0.17  & 0.60$\pm$0.20 \\
HS$_2$   & 3$_{0}$$_{,}$$_{3}$-2$_{0}$$_{,}$$_{2}$, $J$ = 7/2-5/2, $F$ = 4-3  &  47214.73  & 4.5 & 7.07e-7 & 5.89$\pm$0.25   & 0.96$\pm$0.09  & 1.41$\pm$0.50  & 1.44$\pm$0.51 \\
HS$_2$   & 3$_{0}$$_{,}$$_{3}$-2$_{0}$$_{,}$$_{2}$, $J$ = 5/2-3/2, $F$ = 3-2  &  47421.84  & 4.5 & 6.71e-7 & 5.70$\pm$0.18   & 1.00$\pm$0.08  & 1.00$\pm$0.36  & 1.07$\pm$0.32 \\
\hline
\end{tabular}
\label{table:HS2_line_parameters}
\end{table*}

Although sulphur is one of the most abundant species in the Universe, S-bearing molecules are not as abundant as expected in the ISM. In the diffuse ISM, the observed gaseous sulphur accounts \citep[e.g.][]{Howk2006} for its total solar abundance \citep[S/H$\sim$1.5$\times$10$^{-5}$;][]{Asplund2009}. However, in dense molecular clouds and star-forming regions, there is an unexpected paucity of sulphur species. In fact, to reproduce observations in hot corinos and hot cores, a significant sulphur depletion of at least one order of magnitude lower than the solar elemental sulphur abundance must be considered \citep[]{Cernicharo2021a, Esplugues2022, Esplugues2023, Hily-Blant2022, Fuente2023, Cernicharo2024}. In particular, the sum of all the observed gas-phase sulphur abundances constitute only $<$1$\%$ of the expected amount \citep[e.g.][]{Fuente2019} in spite of the numerous sulphur species recently detected by QUIJOTE \citep[e.g.][]{Cernicharo2021a, Cernicharo2021b, Cernicharo2021d, Cabezas2025}. The unknown whereabouts of the remaining S in dense clouds remains an open question. Chemical models predict that, in the dense ISM, atomic S would stick on grains and be mostly hydrogenated forming H$_{{2}}$S \citep{Hatchell1998, Garrod2007, Esplugues2014}, especially at low ($<$20 K) temperatures \citep[]{Druard2012}. However, H$_{{2}}$S has never been detected in interstellar ices. Models and laboratory experiments also suggest that sulphur could be locked into pure S-allotropes (S$_x$) or into hydrogen sulphides (H$_x$S$_y$) \citep[]{Jimenez-Escobar2011, Shingledecker2020}.  This has led to a number of observational searches for these S-species in different star-forming regions \citep{Esplugues2013, Martin-Domenech2016} without success. Only upper limits for H$_2$S$_2$ and HS$_2$ could be derived. In fact, unlike their oxygenated counterparts (H$_2$O$_2$, HO$_2$, and HSO), which have been already detected \citep[]{Bergman2011, Parise2012, Marcelino2023}, H$_2$S$_2$ still remains undetected in space, and HS$_2$ has only been detected in the Horsehead photon-dominated region \citep[]{Fuente2017, Fuente2025}, thus maintaining the mystery of sulphur absence in cold regions.

In this letter, we report the first detection of HS$_2$ in the cold dark cloud TMC-1, providing new insights into the sulphur chemistry in star-forming regions. Observations are described in Sect. \ref{Observations}. We also present and discuss our results in Sects. \ref{section:results} and \ref{section:discussion}, respectively, where we also include the use of chemical models to study the most plausible HS$_2$ formation paths considering gas-phase and surface chemical reactions. We finally summarise our conclusions in Sect. \ref{section:summary}.

\section{Observations}
\label{Observations}

The observations presented in this work are from the ongoing QUIJOTE line survey carried out with the Yebes 40m telescope. A detailed description of the line survey and the data-analysis procedure are provided in \cite{Cernicharo2021c}. QUIJOTE consists of a line spectral survey in the Q band (31.0–50.3 GHz) at the position of the cyanopolyyne peak of TMC-1 ($\alpha$$_{\mathrm{J2000}}$ = 04$^{\mathrm{h}}$:41$^{\mathrm{m}}$:41.9$^{\mathrm{s}}$,  $\delta$$_{\mathrm{J2000}}$ = 25$^{\mathrm{o}}$:41$^{\mathrm{'}}$:27.0$^{\mathrm{''}}$). The survey was performed in several sessions between 2019 and 2024, using new receivers built within the Nanocosmos project\footnote{https://nanocosmos.iff.csic.es/} and installed at the Yebes 40m  telescope \citep[]{Tercero2021}. The Q band receiver consists of two cold high-electron-mobility transistor amplifiers (HEMT) that cover the 31.0-50.3 GHz band with horizontal and vertical polarisations. Fast Fourier transform spectrometers (FFTSs) with 8$\times$2.5 GHz and a spectral resolution of 38.15 kHz ($\sim$0.27 km s$^{-1}$) was also used. The observational mode was frequency-switching with frequency throws of either 10 or 8 MHz. In addition, different central frequencies were used during the runs in order to check that no spurious spectral ghosts were produced in the down-conversion chain. The total observing time on the source for the data taken with the frequency throws of 10 MHz and 8 MHz is 772.6 and 736.6 hours, respectively. Hence, the total observing time on source is 1509.2 hours. The QUIJOTE sensitivity varies between 0.06 mK at 32 GHz and 0.18 mK at 49.5 GHz.  

The intensity scale in antenna temperature ($T$$^{\star}_{\mathrm{A}}$, which is corrected for atmospheric absorption and for antenna ohmic and spillover losses) was calibrated using two absorbers at different temperatures and the atmospheric transmission model \citep[ATM;][]{Cernicharo1985b, Pardo2001}. The absolute calibration uncertainty is 10$\%$.
The data were reduced and processed by using the CLASS package provided within the GILDAS software\footnote{http://www.iram.fr/IRAMFR/GILDAS}, developed by the IRAM Institute.

\begin{figure*}
\centering
\includegraphics[scale=0.63, angle=0]{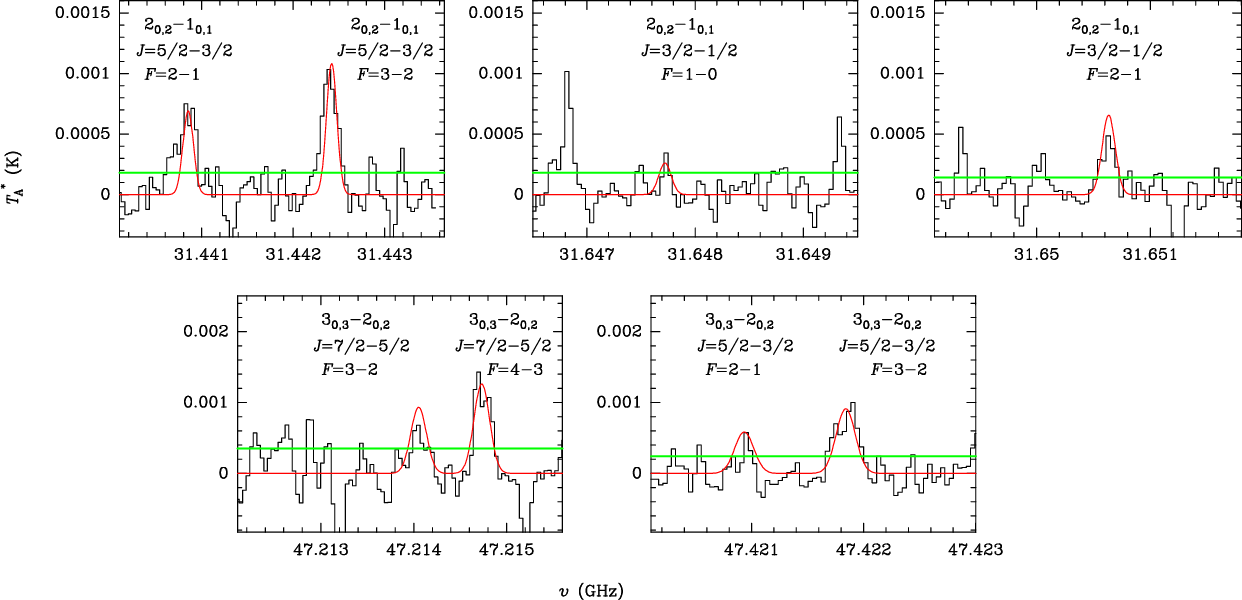}  

\hspace{-1.5cm}
\\
\vspace{-0.5cm}
\caption{Observed lines of HS$_2$ in TMC-1 in the 31.0-50.4 GHz range. Quantum numbers are indicated in each panel. The red line shows the LTE synthetic spectrum from a fit to the observed line profiles. The negative features are created in the folding of the frequency switching data. The horizontal green line indicates the 3 $\sigma$ noise level.}
\label{figure:HS2_lines}
\end{figure*}

\section{Results}
\label{section:results}

In the analysis of the 7 mm line survey of TMC-1, we detected five lines of HS$_2$ corresponding to different hyperfine components of the rotational transitions 2$_{0}$$_{,}$$_{2}$-1$_{0}$$_{,}$$_{1}$ and 3$_{0}$$_{,}$$_{3}$-2$_{0}$$_{,}$$_{2}$ (whose energies are $E$$_{\mathrm{u}}$ = 2.2 and 4.5 K, respectively), using the Cologne Database for Molecular Spectroscopy \citep[CDMS,][]{Muller2001, Muller2005, Endres2016} and the MADEX 
code \citep[]{Cernicharo2012}. This molecule shows fine and hyperfine components due to electron and nuclear spin hyperfine interactions. In particular, we detected four lines of HS$_2$ above 4$\sigma$ (2$_{0}$$_{,}$$_{2}$-1$_{0}$$_{,}$$_{1}$, $J$ = 5/2-3/2, $F$ = 2-1; 2$_{0}$$_{,}$$_{2}$-1$_{0}$$_{,}$$_{1}$, $J$ = 5/2-3/2, $F$ = 3-2; 3$_{0}$$_{,}$$_{3}$-2$_{0}$$_{,}$$_{2}$, $J$ = 7/2-5/2, $F$ = 4-3; and 3$_{0}$$_{,}$$_{3}$-2$_{0}$$_{,}$$_{2}$, $J$ = 5/2-3/2, $F$ = 3-2) and one at the 3.5$\sigma$ level (2$_{0}$$_{,}$$_{2}$-1$_{0}$$_{,}$$_{1}$, $J$ = 3/2-1/2, $F$ = 2-1). The other three hyperfine transitions covered in the Q band (2$_{0}$$_{,}$$_{2}$-1$_{0}$$_{,}$$_{1}$, $J$ = 3/2-1/2, $F$ = 1-0; 3$_{0}$$_{,}$$_{3}$-2$_{0}$$_{,}$$_{2}$, $J$ = 7/2-5/2, $F$ = 3-2; and 3$_{0}$$_{,}$$_{3}$-2$_{0}$$_{,}$$_{2}$, $J$ = 5/2-3/2, $F$ = 2-1) were not detected due to their lower line strength and/or because the noise was significantly higher in that region of the spectrum. The detected lines present antenna temperatures between 0.47 and 1.44 mK, and line-widths ranging from 0.96 to 1.85 km s$^{-1}$, similar to those found for other sulphur species in TMC-1 \citep[]{Marcelino2023, Agundez2025}. In addition, they are not blended with any other spectral feature.  
Table \ref{table:HS2_line_parameters} lists the line parameters obtained from Gaussian fits of the detected lines. Figure \ref{figure:HS2_lines} shows the line profiles (black lines) of HS$_2$ for the transitions 2$_{0}$$_{,}$$_{2}$-1$_{0}$$_{,}$$_{1}$ and 3$_{0}$$_{,}$$_{3}$-2$_{0}$$_{,}$$_{2}$.

In order to calculate the column density of HS$_2$, we used the MADEX code assuming local thermodynamic equilibrium (LTE) due to the lack of collisional rates. An LTE approximation means that most transitions are thermalised to the same temperature
($T$$_{\mathrm{rot}}$ $\sim$ $T$$_{\mathrm{K}}$). If this condition is not met, the temperature might be overestimated, which would produce a variation in the column density. 
In our case, we used a $T$$_{\mathrm{rot}}$ from the transition analysis and find that the considered value is in good agreement with the one obtained for other sulphur molecules in the same region (see below). Nevertheless, it should also be considered that, since the upper energy levels are very low (2.2 and 4.5 K, Table \ref{table:HS2_line_parameters}), the column density is not very sensitive as long as $T$$_{\mathrm{rot}}$ is equal to or higher than 4.5 K.  

The MADEX code takes the different telescope beams and efficiencies into account to obtain antenna temperature intensities for each line. We assumed a diameter size of 80$\arcsec$ \citep[]{Fosse2001}, compatible with the emission size of most of the molecules mapped in TMC-1 \citep[]{Cernicharo2023}. We also considered a line-width of 1.3 km s$^{-1}$ from the average value obtained from the Gaussian fits to the HS$_2$ lines (Table \ref{table:HS2_line_parameters}). We left the rotational temperature and the column density as free parameters. Taken together, the model that best reproduces the line profiles of HS$_2$ is the one with a rotational temperature of 7 K, in good agreement with the results from \cite{Marcelino2023} and \cite{Agundez2025} for the analysis of HSO and CH$_3$CHS, respectively. Figure \ref{figure:HS2_lines} shows the synthetic spectra from the LTE models in red, overlaid with the observed line profiles (black). We derive an HS$_2$ column density of (5.7$\pm$1.1)$\times$10$^{11}$ cm$^{-2}$. The error in the column density is estimated to be 20$\%$. Adopting an H$_2$ column density of 10$^{22}$ cm$^{-2}$ for TMC-1 \citep[]{Cernicharo1987}, we obtain an HS$_2$ abundance relative to H$_2$ of (5.7$\pm$1.1)$\times$10$^{-11}$, which is similar to that obtained by \cite{Fuente2017} in the Horsehead PDR.

\section{Discussion}
\label{section:discussion}

The low temperatures obtained from the LTE models indicate that HS$_2$ emission arises from the cold envelope of TMC-1. Low temperatures have also been observed in TMC-1 for other such sulphur specie as SO \citep[$\sim$4 K,][]{Loison2019} and HSO \citep[$\sim$4.5 K,][]{Marcelino2023}, which suggests a common origin. Particularly interesting is the comparison between HS$_2$ and its oxygenated counterparts HO$_2$ and HSO since sulphur belongs to the chalcogens, along with oxygen. Their similar electron configuration could indicate a similar chemistry; nevertheless, their abundances are quite different. In particular, the abundance of HSO relative to H$_2$ is 7$\times$10$^{-12}$ \citep[]{Marcelino2023} is almost one order of magnitude lower than that of HS$_2$ in TMC-1. 
This trend of low HSO abundance with respect to HS$_2$ is maintained throughout the time evolution of the cloud, according to the chemical model shown in Fig. \ref{figure:S_molecule_models} (Appendix \ref{tables_figures}), which was obtained using the public Nautilus chemical code\footnote{https://forge.oasu.u-bordeaux.fr/LAB/astrochem-tools/pnautilus}. Nautilus is a three-phase model \citep[gas, grain surface, and grain mantle,][]{Ruaud2016} (see Appendix \ref{Nautilus_appendix} for more details). Figure \ref{figure:S_molecule_models} also shows the time evolution of the fractional abundances for H$_2$S, S$_2$, and HO$_2$, which has been detected only in $\rho$ Ophiuchi A with an abundance of $\sim$10$^{-10}$ \citep[]{Parise2012}. In Fig. \ref{figure:S_molecule_models}, H$_2$S and HO$_2$ have the highest and lowest abundances for $t$$>$10$^3$ yr, respectively, with differences among them of several orders of magnitude. We also observe that HO$_2$ and HSO follow a similar trend along the entire cloud evolution.

\subsection{HS$_2$ formation and destruction}
\label{chemical_rates}

Despite the numerous observations, laboratory experiments, and chemical modelling studies performed \citep[e.g.,][]{Hatchell1998, Ruffle1999, Wakelam2008, Vidal2017, Esplugues2023}, sulphur chemistry in the ISM, especially in dark clouds, is one of the least understood. All of these effotts have tried to shed light on the so-called sulphur depletion problem; however, results have been unsuccessful in predicting fractional abundances of species, such as CS, falling short by up to two orders of magnitude compared to astronomical observations. Chemical models also predict that, in the dense ISM, atomic sulphur would stick on grains and be mostly hydrogenated, or be locked into solid species, such as H$_2$S$_2$, CS$_2$, and S$_8$ \citep[]{Palumbo1997, Shingledecker2020, Cazaux2022}. However, the lack of observations of these species makes understanding sulphur chemistry difficult. In this way, the detection of new sulphur species in dark clouds, as well as the study of their formation mechanisms, is significantly valued.   
 
In this regard, we carried out chemical modelling calculations using the Nautilus code to study the main formation and destruction paths of HS$_2$. Figure \ref{figure:HS2_rates} (Appendix \ref{tables_figures}) shows the chemical rates for the most dominant reactions forming and destroying HS$_2$. In order to evaluate the density effect on the chemical rates, two different models were considered: one model with a cloud density of $n$$_{\mathrm{H}}$=10$^4$ cm$^{-3}$ (solid lines) and another model with $n$$_{\mathrm{H}}$=10$^5$ cm$^{-3}$ (dashed lines). We observe that the effect of varying the cloud density mainly affects the chemical rates at early evolutionary times ($t$$\lesssim$10$^5$ yr), with differences of several orders of magnitude between both models. Nonetheless, we note that these differences become smaller as the cloud evolves.

We find that HS$_2$ is mainly formed through the dissociative recombination of the ions H$_3$S$^{+}_{\mathrm{2}}$ and H$_2$S$^{+}_{\mathrm{2}}$, following

\vspace{-0.3cm}

\begin{equation}
\mathrm{H_3S_2^+ + e^- \rightarrow H_2 + HS_2} 
\label{equation:1}
\end{equation}

\vspace{-0.5cm}

\begin{equation}
\mathrm{H_2S_2^+ + e^- \rightarrow H + HS_2,} 
\label{equation:2}
\end{equation}

\noindent which are mainly formed in turn from the reactions: 

\vspace{-0.3cm}

\begin{equation}
\mathrm{H_2 + HS_2^+ \rightarrow H_3S_2^+} 
\label{equation:3}
\end{equation}

\vspace{-0.5cm}

\begin{equation}
\mathrm{HSSH + H^+ \rightarrow H + H_2S_2^+,} 
\label{equation:4}
\end{equation}

\noindent respectively. In particular, we find that reaction \ref{equation:1} is the dominant reaction forming HS$_2$ during $t$$\lesssim$10$^5$ yr, and that for longer times reaction \ref{equation:2} also becomes important. By contrast, we do not find any significant contribution from the surface reactions in the formation of HS$_2$. Nevertheless, we should notice that some relevant surface reactions for HS$_2$, such as 

\vspace{-0.5cm}

\begin{equation}
\mathrm{JS_2 + JH \rightarrow JHS_2,} 
\end{equation}

\noindent are not included in the chemical network. While other reactions, such as  

\vspace{-0.5cm}

\begin{equation}
\mathrm{JH_2 + JS_2 \rightarrow JHSSH,} 
\end{equation}

\noindent are considered as barrierless reactions. This significantly reduces the S$_2$ abundance on the grains and, consequently, partially deactivates that chemistry. All this highlights the poor knowledge we currently have on HS$_2$ chemistry on grain surfaces.

Regarding its destruction, Figure \ref{figure:HS2_rates} shows that HS$_2$ is mainly destroyed at early times ($t$$<$10$^4$ yr) by

\vspace{-0.3cm}

\begin{equation}
\mathrm{HS_2 + H^+ \rightarrow H + HS_2^+}, 
\end{equation}

\noindent while for a more evolved cloud (10$^4$$\lesssim$$t$$\lesssim$10$^6$ yr), the ions H$^{+}_{\mathrm{3}}$, HCO$^+$, and H$_3$O$^+$ play an important role in destroying HS$_2$. In any case, there are also many reactions between HS$_2$ and neutral atoms, such as H, C, or O, which are neither included in the network nor studied in detail, and which could significantly alter the theoretical HS$_2$ abundances obtained in the models. In fact, only a few analyses on the chemistry of HS$_2$ (only at high temperatures) are found in the literature \citep[e.g.,][]{Sendt2002}.

\subsection{Model vs observations}
\label{model_observations}

In order to theoretically reproduce the HS$_2$ abundance obtained in TMC-1, and evaluate the influence of the physical parameters on HS$_2$, we ran several models (Table \ref{table:HS2_models}, Appendix \ref{tables_figures}) varying density, temperature, and initial sulphur abundance. All the models were run considering $A$$_{\mathrm{V}}$ = 15 mag, which was derived from visual extinction maps from Herschel data \citep[]{Navarro-Almaida2021}, and a cosmic-ray ionisation rate of $\zeta$=1.3$\times$10$^{-17}$ s$^{-1}$, according to the value derived by \cite{Navarro-Almaida2021} in TMC-1 by comparing spectroscopic observations with theoretical results. Figure \ref{figure:model_HS2_observations} (Appendix \ref{tables_figures}) shows our results. We find that the observations are reproduced by all the models considering an initial sulphur abundance of 1.5$\times$10$^{-6}$. In general, we observe that the more depleted the sulphur is, the lower the HS$_2$ abundance throughout the cloud's evolution. In particular, decreasing the initial sulphur abundance by one order of magnitude leads to a decrease in HS$_2$ abundance by about 2-3 orders of magnitude. The density also has a key role in HS$_2$ abundance, especially at $t$$<$10$^5$ yr, where we find that increasing $n$$_{\mathrm{H}}$ by one order of magnitude leads to significantly higher abundances of HS$_2$ owing to the higher collision rates of reactions \ref{equation:1}-\ref{equation:4}.

In any case, our models reproduce the observations of HS$_2$ when $S$$^{+}_{\mathrm{init}}$=1.5$\times$10$^{-6}$, for an evolutionary time of $t$$\gtrsim$10$^4$ yr if the cloud density is $n$$_{\mathrm{H}}$=10$^{4}$ cm$^{-3}$, or at $t$$>$10$^3$ yr if $n$$_{\mathrm{H}}$=10$^{5}$ cm$^{-3}$.

\section{Conclusions}
\label{section:summary}

We report the first detection of HS$_2$ in a cold dark cloud (TMC-1), thanks to the highly sensitive QUIJOTE Q-band spectral line survey. This is the first observation of a chemical species containing more than one sulphur atom in this type of source. Our analysis reveals that its molecular emission arises from the very cold envelope of TMC-1, with an abundance similar to that found in the Horsehead PDR, but almost one order of magnitude higher than that for its oxygenated counterpart HSO in TMC-1. A chemical analysis also shows that gas-phase chemical reactions, in particular dissociative recombination reactions, play a major role forming HS$_2$. This molecules is very sensitive to the cloud density and to the sulphur depletion degree.

\begin{acknowledgements}

This work was based on observations carried out with the Yebes 40m telescope (projects 19A003, 20A014, 20D023, 21A011, 21D005, and 23A024). Yebes 40m telescope is operated by the Spanish Geographic Institute (IGN, Ministerio de Transportes y Movilidad Sostenible). We acknowledge funding support from Spanish Ministerio de Ciencia e Innovación through grants PID2022-137980NB-I00, PID2022-136525NA-I00 and PID2023-147545NB-I00. G. E. acknowledges the ERC project SUL4LIFE (grant agreement No101096293) from European Union.
 G. M. acknowledges the support of the grant RYC2022-035442-I funded by MICIU/AEI/10.13039/501100011033 and ESF+. G. M. also received support from project 20245AT016 (Proyectos Intramurales CSIC). We also thank the anonymous referee for valuable comments that improved the manuscript.

\end{acknowledgements}

\bibliographystyle{aa}
\bibliography{biblio}

\begin{appendix}

\section{Nautilus chemical code}
\label{Nautilus_appendix}

The Nautilus chemical code \citep[]{Ruaud2016} solves the kinetic equations for both the gas phase species and the grain surface species, and computes the time evolution of chemical abundances.  
Gas-phase chemical reactions included in Nautilus are bimolecular reactions (neutral-neutral and ion-neutral reactions), direct cosmic-ray ionisation or dissociation, ionisation and dissociation by UV photons, ionisation and dissociation produced by photons induced by cosmic-ray interactions with the medium \citep[]{Prasad1983}, and electron recombinations. Regarding chemical grain surface processes, the code includes sputtering of grains by cosmic ray particles (CR sputtering), and non-thermal desorption (distinguishing between three types of desorption: photo-desorption, chemical desorption, and cosmic-ray heating). These three types of desorption are only allowed to occur for surface species, while CR sputtering can take place for both surface and mantle species. The chemical network used in Nautilus is based on the KInetic Database for Astrochemistry (KIDA\footnote{https://kida.astrochem-tools.org/}). In particular, it is the chemical network kida.uva.2022, which contains the updates presented in \cite{Wakelam2024} and new non-thermal desorption mechanisms, such as cosmic-ray sputtering, as described in \cite{Wakelam2021}. Nautilus is run considering the initial abundances shown in Table \ref{table:abundances_Nautilus}.

\begin{table}[h!]
\caption{Abundances with respect to total hydrogen nuclei considered in the chemical code Nautilus.
}                 
\centering  
\begin{tabular}{l l l l l}     
\hline\hline       
Species &  Abundance   & Reference                       \\ 
\hline 
He      & 9.0$\times$10$^{-2}$   &  (1) \\
O       & 2.4$\times$10$^{-4}$   &  (2) \\
Si$^+$  & 8.0$\times$10$^{-9}$   &  (3) \\ 
Fe$^+$  & 3.0$\times$10$^{-9}$   &  (3) \\ 
S$^+$   & 1.5$\times$10$^{-6}$, 1.5$\times$10$^{-7}$   &  (4)  \\
Na$^+$  & 2.0$\times$10$^{-9}$   &  (3)  \\
Mg$^+$  & 7.0$\times$10$^{-9}$   &  (3)  \\
P$^+$   & 2.0$\times$10$^{-10}$  &  (3)  \\
Cl$^+$  & 1.0$\times$10$^{-9}$   &  (3)   \\
F$^+$   & 6.7$\times$10$^{-9}$   &  (5)  \\
N       & 6.2$\times$10$^{-5}$   &  (6) \\
C$^+$   & 1.7$\times$10$^{-4}$   &  (7) \\
H$_2$   & 0.5                &  (8) \\
\hline 
\label{table:abundances_Nautilus}                 
\end{tabular}
\tablefoot{
References: (1) \cite{Asplund2009} and \cite{Wakelam2008}. (2) \cite{Wakelam2008} and \cite{Navarro-Almaida2021}. (3) As in the low-metal abundance case from \cite{Graedel1982} and \cite{Morton1974}. (4) We considered a depletion factor of 10-100 with respect to the sulphur cosmic elemental abundance of 1.5×10$^{-5}$ to take into account the recent sulphur depletion results from \cite{Esplugues2022}, \cite{Esplugues2023}, and \cite{Fuente2023}. (5) \cite{Neufeld2005}. (6) \cite{Navarro-Almaida2021} and \cite{Jimenez-Serra2018}. (7) \cite{Navarro-Almaida2021}. (8) \cite{Wakelam2021}.\\
} 
\\
\end{table}

\section{Tables and Figures}
\label{tables_figures}

\begin{figure}
\centering
\includegraphics[scale=0.44, angle=0]{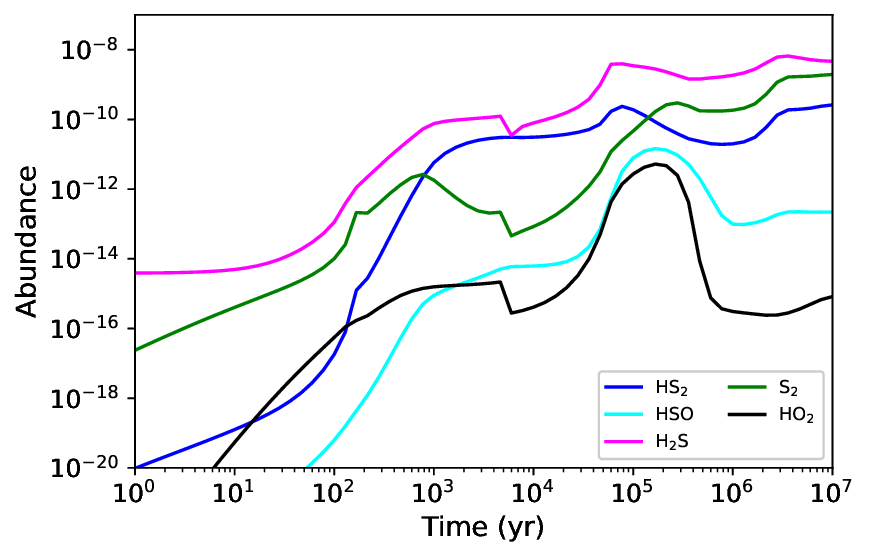}  
\hspace{-0.5cm}
\vspace{-0.2cm}
\\
\caption{Evolution of fractional abundances ($n$$_{\mathrm{x}}$/$n$$_{\mathrm{H}}$) of HS$_2$ (blue), HSO (cyan), H$_2$S (magenta), S$_2$ (green), and HO$_2$ (black) as a function of time for an initial sulphur abundance of $S$$^{+}_{\mathrm{init}}$=1.5$\times$10$^{-6}$, a hydrogen number density of $n$$_{\mathrm{H}}$=10$^5$ cm$^{-3}$, $T$$_{\mathrm{gas}}$=5 K, and a CR ionisation rate of $\zeta$=1.3$\times$10$^{-17}$ s$^{-1}$.}
\label{figure:S_molecule_models}
\end{figure}

\begin{figure}
\centering
\includegraphics[scale=0.41, angle=0]{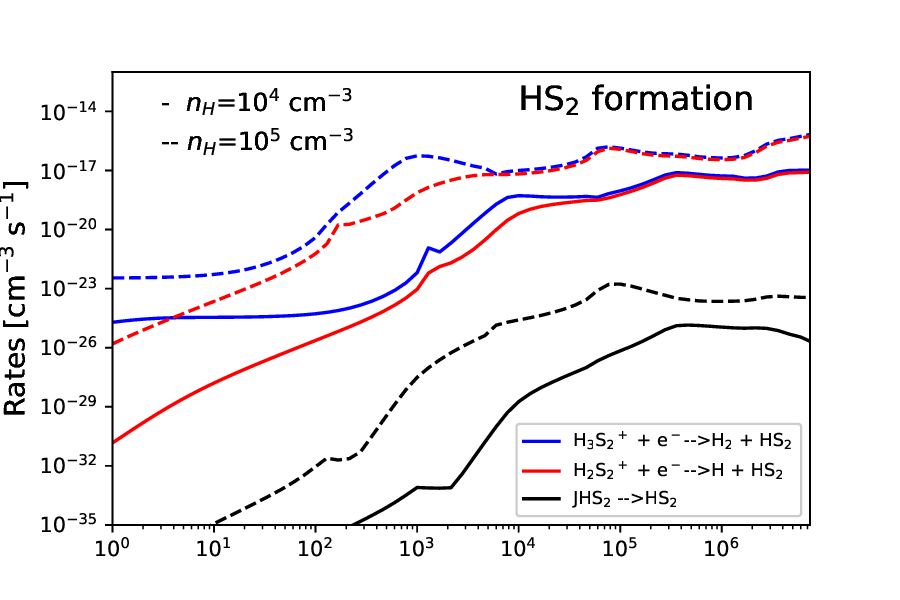} \vspace{0.1cm}\\
\vspace{-0.7cm}
\includegraphics[scale=0.41, angle=0]{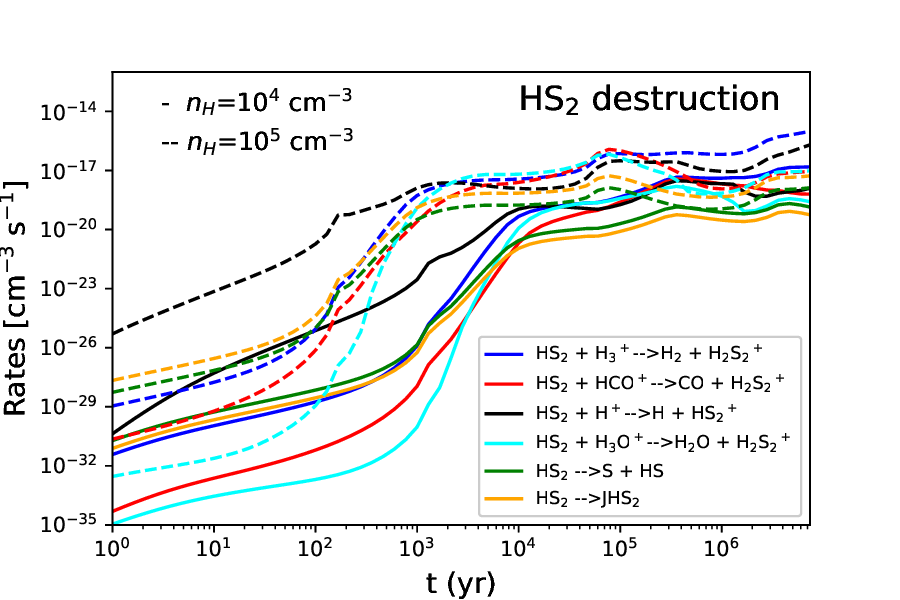}\\
\caption{Main chemical reaction rates forming (top) and destroying (bottom) HS$_2$. JX means solid X.}
\label{figure:HS2_rates}
\end{figure}

\begin{table}
\centering
\caption{Adopted model parameters.}
\begin{tabular}{llll}
\hline 
\hline
Model    & $n$$_{\mathrm{H}}$   & $T$$_{\mathrm{g}}$ &  $S$$^{+}_{\mathrm{init}}$   \\
         & cm$^{-3}$            & (K)                &                              \\
\hline
\hline 
1   & 10$^{4}$ &  5  & 1.5$\times$10$^{-6}$  \\
2   & 10$^{5}$ &  5  & 1.5$\times$10$^{-6}$  \\
3   & 10$^{4}$ &  10 & 1.5$\times$10$^{-6}$  \\
4   & 10$^{4}$ &  5  & 1.5$\times$10$^{-7}$ \\
\hline
\end{tabular}
\label{table:HS2_models}
\end{table}

\begin{figure}
\centering
\includegraphics[scale=0.44, angle=0]{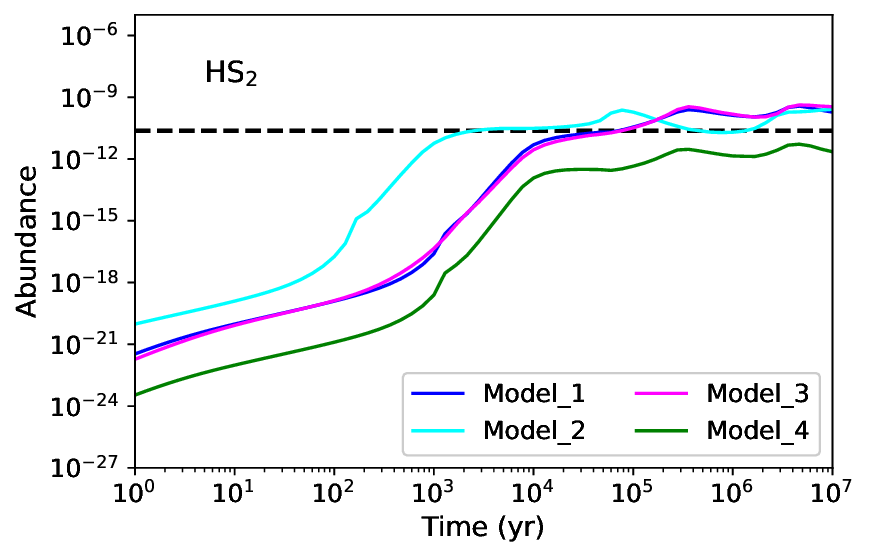}  
\hspace{-0.5cm}
\vspace{-0.2cm}
\\
\caption{Comparison of the observational HS$_2$ abundance in TMC-1 with the fractional abundances obtained with Models 1-4. Dashed black line represents the observed HS$_2$ abundance.}
\label{figure:model_HS2_observations}
\end{figure}

\end{appendix}

\end{document}